\documentclass[aps,prb,eps,superscriptaddress,twocolumn,10pt]{revtex4}
\usepackage{graphicx}
\usepackage{lipsum}
\usepackage{amssymb}
\usepackage{amsmath}
\usepackage{color}
\newcommand\correspondingauthor{\thanks{t.sahadasgupta@gmail.com}}

\definecolor{ojcolor}{cmyk}{0.63, 0.33, 0, 0.26}

\definecolor{ojcolor}{cmyk}{0.8, 0, 0, 0.26}

\begin{document}
\title{Electronic and Magnetic State of LaMnO$_3$ Epitaxially Strained on SrTiO$_3$: Effect of Local Correlation and
Non-local Exchange}
\author{Hrishit Banerjee}
\affiliation{Insitut f\"{u}r Theoretische Physik - Computational Physics, TU Graz, Petersga{\ss}e 16, Graz, 8010, Austria.}
\author{Oleg Janson}
\affiliation{Institute for Theoretical Solid State Physics, IFW Dresden, Helmholtzstra{\ss}e 20, 01069 Dresden, Germany}
\affiliation{Institut f\"{u}r Festk{\"{o}}rperphysik, TU Wien, Wiedner Hauptstra{\ss}e 8-10, Vienna, Austria}
\author{Karsten Held}
\affiliation{Institut f\"{u}r Festk{\"{o}}rperphysik, TU Wien, Wiedner Hauptstra{\ss}e 8-10, Vienna, Austria}
\author{Tanusri Saha-Dasgupta\correspondingauthor}
\email{t.sahadasgupta@gmail.com}
\affiliation{S. N. Bose National Centre for Basic Sciences, JD Block, Sector III, Salt Lake, Kolkata, West Bengal 700106, India.}
\affiliation{Indian Association for the Cultivation of Science, 2A \& 2B, Raja Subodh Chandra Mallick Rd, Jadavpur, Kolkata, West Bengal 700032, India.}
\begin{abstract}
  Motivated by the puzzling report of the observation of a ferromagnetic insulating state in LaMnO$_3$/SrTiO$_3$ heterostructures, we calculate the electronic and magnetic state of LaMnO$_3$, coherently matched to a SrTiO$_3$ square substrate within a
  ``strained-bulk'' geometry. We employ three different density functional theory based computational approaches: (a)
  density functional theory (DFT)
  supplemented with Hubbard $U$ (DFT+$U$), (b) DFT + dynamical mean field theory (DMFT), and (c) a hybrid functional treatment of the
  exchange-correlation functional. While the first two approaches include  local correlations and exchange at Mn sites,
  treated in a static and dynamic manner, respectively, the last one takes into account the effect of non-local exchange
  at all sites. We find in all three approaches that the compressive strain induced by the square substrate
  of SrTiO$_3$ turns LaMnO$_3$ from an antiferromagnet with sizable orbital polarization to a  ferromagnet with suppressed Jahn-Teller distortion
  in agreement with experiment. However, while both DFT+$U$ and DFT+DMFT provide
  a metallic solution,  only the hybrid calculations result in an insulating solution, as observed in experiment.
  This insulating behavior is found to originate from an electronic charge disproportionation.
  Our conclusions remain valid when we investigate  LaMnO$_3$/SrTiO$_3$ within the experimental set-up of a superlattice geometry using DFT+$U$
  and hybrid calculations.
\end{abstract}

\maketitle

\section{Introduction}
In recent times the study of interfaces formed between perovskite oxides has made a deep impact on the community engaged in both theoretical and experimental condensed matter research. The presence of a highly conducting 2-dimensional electron gas (2DEG) at the interface between oxide insulators
grown along the [001] direction, has opened a road  to many different prospective device applications. Interfaces have
been formed between band insulators, like between LaAlO$_3$ (LAO) and SrTiO$_3$ (STO),\cite{lao-sto} generating significant excitement at the 2DEG interface.\cite{sc-fm}
This excitement has been propagated further in terms of synthesis of interfaces between Mott insulators and band insulators,
as in case of GdTiO$_3$ (GTO) and SrTiO$_3$,\cite{gto-sto} or LaTiO$_3$ and SrTiO$_3$.\cite{lao-sto} The behavior of
GTO/STO interface has been
found to be qualitatively different from that of the LAO/STO interface in terms of the absence of a critical thickness for
metallicity; and the carrier density of the 2DEG being in perfect agreement with that expected from polar charge.\cite{stemmer} The qualitatively
different behavior of the two systems has been rationalized by the fact that the gap in GdTiO$_3$ being a Mott gap arises
within the Ti $d$ manifold, while that in LaAlO$_3$ arises between filled O $p$ bands and empty Al bands, thereby
influencing the band alignment in a qualitative manner.\cite{banerjee-ch6}

Given the dissimilar behavior of band and Mott insulators in oxide heterostructures, it
is curious to ask what happens if an insulator such as LaMnO$_3$ (LMO)
featuring a cooperative Jahn-Teller (JT) distortion of MnO$_6$ octahedra along with the strong onsite repulsion, is brought in contact with the band insulator
SrTiO$_3$. The influence of structural distortions is expected to be qualitative in this case,
though LAO, GTO and LMO all belong to the same polar family. Indeed LMO/STO heterostructures
have been synthesized and probed experimentally to characterize
the nature of the interfaces, formed between LMO and STO. Varied nature of magnetic and electronic behavior of the LMO/STO interfaces have been reported,
depending on the relative thickness of LMO and STO and their geometry.\cite{barriocanal-ch6,garcia-ch6,choi-ch6,liu2-ch6,sumilan-ch6,wang2015,oor} 
Among all, the most intriguing is the suggestion of ferromagnetic insulating behavior, which has been reported for LMO/STO 
superlattices when LMO and STO have comparable thicknesses,\cite{oor} as well as in thin-film/substrate geometries.\cite{sumilan-ch6} This is counter-intuitive, since ferromagnetism is commonly associated with metallicity and antiferromagnetism
is typical for insulators. Some attempts have been made to justify the observed coexistence of ferromagnetism and insulating nature. One of them involved Monte-Carlo simulation\cite{oor} of a double-exchange model, with orbital polarization to
explain the behavior. However, experimental investigation shows significant suppression of the JT distortion\cite{oor} in
the superlattice geometry showing ferromagnetic insulating behavior, and thus the orbital polarization
is also expected to be suppressed. The other one relies on the concept of electronic phase separation leading to
nucleation of
metallic nanoscale ferromagnetic islands embedded in an insulating antiferromagnetic matrix, which gives rise to both ferromagnetic
signal and insulating resistivity.\cite{sumilan-ch6} For this scenario to be valid, the polar charge created at the interface, must reside at the LMO side, resulting in doping of LMO. The direction of charge transfer, however has not been
established so far. The observed coexistence of ferromagnetism and
insulating nature at LMO/STO interface thus remains a puzzle. Is it  intrinsic or due to extrinsic reasons such as deviations from stoichiometry in the heterostructure, the presence of defects that
trap the free carriers, or inhomogeneity of the samples? 

Investigations using density functional theory (DFT)  supplemented with Hubbard $U$ (DFT+$U$) on strained LMO 
corresponding to that of STO showed \cite{spaldin-ch6} a  suppression of the JT distortion and a ferromagnetic 
ground state, which is albeit metallic rather than insulating as claimed in experiments.
In a further DFT+$U$ study of LMO strained to STO,\cite{hou-ch6} the structural relaxation allowing for symmetry lowering to monoclinic structure,
and resultant antiferro-orbital ordering between symmetry inequivalent Mn atoms was used to explain the ferromagnetic insulating
behavior of LMO. However, such symmetry lowering may be difficult to be accommodated within a heterostructure geometry, where
LMO might be sandwiched between the layers of STO, thus being constrained from both top and bottom and unable to deviate from the cubic
group of symmetry.

In view of the above, we revisit the problem considering three different computational approaches on bulk LMO with its in-plane lattice
constants constrained to that corresponding to a square substrate geometry of STO. We consider
the general framework of DFT, which is expected to capture the structural changes that happen upon epitaxial straining of LMO
correctly. To take into account the strong correlation effect at transition metal (TM) site, which is known to be essential in
proper description of magnetic
and electronic ground states of manganites, we follow three different methods: (i) The static treatment of correlations including an orbital-dependent potential that is parametrized in terms of Hubbard parameter $U$ and Hund parameter $J_H$ within the DFT+$U$ formulation,\cite{dudarev-ch6} 
as followed in previous literature.\cite{spaldin-ch6,hou-ch6} (ii) Hybrid functional as implemented by Heyd-Scuseria-Ernzerhof (HSE) \cite{hse-ch6} in which a portion of the exact 
non-local Hartree-Fock (HF) exchange is mixed with the complementary DFT in local (LDA) or semilocal (GGA) approximated exchange. As opposed to the ``+$U$'' formulation where the improved treatment of exchange
effects is limited to the partially filled TM sites, the hybrid functional approach uses an orbital-dependent functional acting
on all states, extended as well as localized. It  thus has impact on both TM sites and O sites. This may become
important in strongly covalent systems as manganites. Indeed, as argued
in Ref.\onlinecite{cesare} the bulk properties of LMO is reproduced better in hybrid calculation, compared
to DFT+$U$. (iii) DFT+ Dynamical Mean Field Theory (DMFT)\cite{LDADMFT} in order to probe the effect of dynamical correlation as well as that of temperature. In order to minimize the computational effort, the DMFT calculations were carried out employing the Mn $d$ only low-energy Hamiltonian, consisting
of five orbitals per Mn site, derived out of DFT in the maximally localized Wannier function basis.

We find a ferromagnetic
ground state in all three approaches, driven by the marked reduction of orthorhombic distortion in the optimized LMO structure
when epitaxially strained to the square substrate of STO, resulting in a strong suppression of the JT distortion.
The suppression of the JT distortion and modification of the octahedral rotation, as captured in our study, is in agreement with structural characterization of LMO/STO superlattices,\cite{oor} stressing once again the accuracy of DFT in addressing
the structural properties. Although the three methods agree on the magnetic state of strained LMO, the DFT+$U$ and DFT+DMFT resulted in
metallic solutions, while the treatment of correlation effect within hybrid functional resulted in an insulating solution. This surprising
result of ferromagnetic insulating solution within hybrid calculations, was traced to originate from electronically driven charge disproportion (CD) within the Mn sublattice
that arises due to a strain-driven enhanced covalency between Mn and O. We note that in the hybrid approach, as opposed to
both, DFT+$U$ and DFT+DMFT, the exact exchange is calculated for all the orbitals, not only for the TM sites. This, in turn, presumably highlight
the Mn-O covalency effect, and thus the importance of correlation effects on the O $p$ states, which are considered uncorrelated or
with no self-energy in the conventional DFT+$U$ and DFT+DMFT set-up.

Finally, in order to further probe the effect of the heterostructure geometry, as in the experimental set-up, we compare 
the results of DFT+$U$ and hybrid calculations for (LMO)$_{4.5}$/(STO)$_{4.5}$ 
with two symmetric n-type interfaces in superlattice geometry, which takes into account
the presence of STO in an explicit manner. The calculations on heterostructure geometry
confirm the ferromagnetic insulating result for hybrid and ferromagnetic
metallic state for DFT+$U$ approach, making our conclusion to remain valid even in experimentally relevant geometry.
We hope that our extensive theoretical study will regenerate interest in the curious case of ferromagnetic insulating
state of LMO/STO interface, in terms of better characterization of
the samples with detailed knowledge on oxygen vacancies, defects and inhomogeneities on one hand, and a more
complete many-body treatment taking into account the oxygens explicitly within a charge self-consistent scheme
on the other.

\section{Computational Details}

Our DFT calculations were carried out in a plane wave basis  with projector-augmented wave (PAW) potentials\cite{blochl-ch6} as implemented in the Vienna Ab-initio Simulation Package
(VASP).\cite{kresse-ch6, kresse01-ch6} The DFT
exchange-correlation functional was chosen to be that given by generalized gradient approximation (GGA), implemented
following the Perdew Burke Ernzerhof (PBE) prescription.\cite{pbe-ch6}
For ionic relaxations, internal positions of the atoms were allowed to relax until the forces became less than 0.005 eV/$\AA$. An energy cutoff of 550 eV, and 5 $\times$ 5 $\times$ 3 Monkhorst–Pack k-points mesh were found to provide a good convergence of the total energy in self-consistent field calculations. The
plane wave cutoff and the k-point mesh have been checked
for convergence of the  obtained results. PAW-PBE potentials with highest available energy cutoff of 220eV for La (11 valence $e^{-}$), 270eV for Mn (13 valence $e^{-}$) , and 400eV for O (6 valence $e^{-}$) have been used. 

The DFT+$U$ calculations were carried out in form of GGA+$U$. The value of $U$ at the Mn sites in the GGA+$U$ scheme was varied from 2eV to 8eV; and
a $U$ value of 3.5eV was found to be adequate to reproduce the experimentally observed insulating A-AFM nature of bulk unstrained LaMnO$_3$.
The Hund's coupling parameter $J_H$ was chosen be 0.9 eV.

The functional used in hybrid calculation can be mathematically expressed as,
\begin{align}
E_{XC}^{HSE}(\omega)= & \alpha E_X^{HF,SR}(\omega)+(1-\alpha)E_X^{PBE,SR}(\omega) \notag\\ & + E_X^{PBE,LR}(\omega)+E_C^{PBE}
\end{align}
where $\alpha$ is the mixing parameter and $\omega$ is an adjustable parameter controlling the short-rangeness of the interaction. Here $E_X^{HF,SR}$ denotes the short range HF exchange functional, $E_X^{PBE,SR}$ denotes the short range PBE exchange functional, $E_X^{PBE,LR}$ indicates the long range PBE exchange functional, and $E_C^{PBE}$ refers to the correlation functional as given by PBE. The standard value of $\omega$=0.2 (referred to as HSE06) along with varying values of $\alpha$  of 0.15, 0.20, 0.25, and 0.30 were used in our calculations. The influence of  the mixing factor, $\alpha$, in
hybrid  functionals has been systematically studied for 3d$^{0}$ - 3d$^{8}$ transition-metal perovskites LaMO$_3$ (M = Sc-Cu) by
He and Franchini,\cite{cesare} which concludes that for LMO the choice of $\alpha$ = 0.15 reproduces the experimental band gaps,
magnetic moments and exchanges best. However the calculations by He and Franchini\cite{cesare} were carried with potentials with
lower cutoffs of 137eV for La (9 valence $e^{-}$), 270eV for Mn (7 valence $e^{-}$) , and 283eV for O (6 valence $e^{-}$), available at that time, with a maximum cutoff energy of 300eV. Repeating the calculations using newer potentials with higher cutoff available
now, as mentioned previously, we find the agreement with experimental results to be best for $\alpha$ = 0.25, which is the standard hybrid functional value.
In the Appendix, we show the MARE (Mean Absolute Relative Error) for the band gaps and the magnetic moments, confirming better agreement
with experimental band gap and magnetic moments for $\alpha$ = 0.25 compared to previously suggested\cite{cesare} $\alpha$ = 0.15.
All calculations reported in the following, were thus carried out for a choice of $\alpha$ value of 0.25.

The starting point of our DFT+DMFT calculations, were GGA calculations
performed in the full potential augmented plane wave basis as implemented in
\textsc{wien2k}.\cite{wien2k18} We used the largest possible muffin-tin radii and the
basis set plane-wave cutoff as defined by
${R_{\text{min}}\!\cdot\!K_{\text{max}}}$\,=\,7, where $R_{\text{min}}$ is the
muffin-tin radius of oxygen atoms. The consistency between plane-wave basis and augmented plane wave basis results has been
cross-checked. The Mn $d$ band structure of a nonmagnetic GGA
calculation is split into the $t_{2g}$ and $e_g$ manifolds, comprising 12 and 8
bands, respectively, as expected for a unit cell with with four Mn atoms and
five orbitals per Mn atom. In a nonmagnetic DFT calculation, the Fermi level
crosses the $t_{2g}$ manifold, the $e_g$ states are empty. 
The maximally localized Wannier functions \cite{wannier90} of these  five  Mn $d$ DFT orbitals were used as an input for DMFT and calculated with the 
\textsc{wien2wannier} interface.~\cite{wien2wannier} As common for multisite DMFT calculations, each of the four Mn
atoms in the unit cell was treated as an independent five-orbital DMFT impurity
problem. Note that in contrast to earlier DMFT studies,\cite{yamasaki06,
yang07, pavarini10} we treated $t_{2g}$ and $e_g$ electrons on the same
footing.  Previous DFT+DMFT calculations for bulk LaMnO$_3$ employed the
intra-orbital Coulomb repulsion $U$\,=\,5\,eV in Refs.~\cite{yamasaki06,
yang07} or $U$\,=\,4--7\,eV in Ref.~\cite{pavarini10} with the Hund's exchange
of $J$\,=\,0.75 eV. Using these values as a starting point, we varied these
parameters over a reasonable range to gain better understanding of their
influence onto the physics of our five-orbital model.

The auxiliary Anderson impurity problems were solved using the
continuous-time quantum Monte Carlo algorithm in the hybridization expansion
(CT-HYB)~\cite{werner06} as implemented in
\textsc{w2dynamics}~\cite{w2dynamics}. Since this algorithm scales
exponentially with the number of orbitals, solving a five-orbital problem turns
into a cumbersome task. To keep the impurity problem numerically tractable, we
performed a non-charge-self-consistent calculation~\cite{footnotedmft} and employed the
recently developed superstate sampling, where impurity eigenstates are grouped
and each group is sampled individually.\cite{kowalski19} The fully localized
limit\cite{anisimov93} was used as the double counting correction.  We used the
rotationally invariant Kanamori interaction~\cite{kanamori63} with the
intra-orbital interaction $U'\!=\!U\!-\!2J$, as commonly employed in DFT+DMFT
calculations. The resulting Hamiltonian accounts for the spin flip and pair
hopping terms, but neglects the different spatial extent of $t_{2g}$ and $e_g$
Wannier functions.

\section{Results}

\subsection{LMO Epitaxially Strained on Square Substrate of STO}

\subsubsection{Crystal Structure}

Bulk unstrained LMO grow in the orthorhombic $Pbnm$ crystal structure, which is derived out of the cubic structure
by expansion of the unit cell to a $\sqrt{2}\times \sqrt{2}\times 2$ perovskite supercell, resulting into four formula
units in the cell, that can accommodate the GdFeO$_3$ type rotation and tilt of the oxygen octahedra as well as JT
distortion. In order to mimic the effect of epitaxial strain, as in Refs.\onlinecite{spaldin-ch6,hou-ch6}, 
we performed ``strained-bulk'' calculations, in which the structural parameters ($c$ lattice parameter,
ionic positions) of the $\sqrt{2}\times \sqrt{2}\times 2$ perovskite supercells were 
optimized subject to the constraint that the two in-plane lattice vectors which define the epitaxial 
substrate, were fixed to produce the specified square lattice of dimensions $\sqrt{2} \times a_{c}$,
where $a_c$ cubic lattice parameter corresponding to the substrate.

\begin{figure}
\centering
\includegraphics[width=\columnwidth]{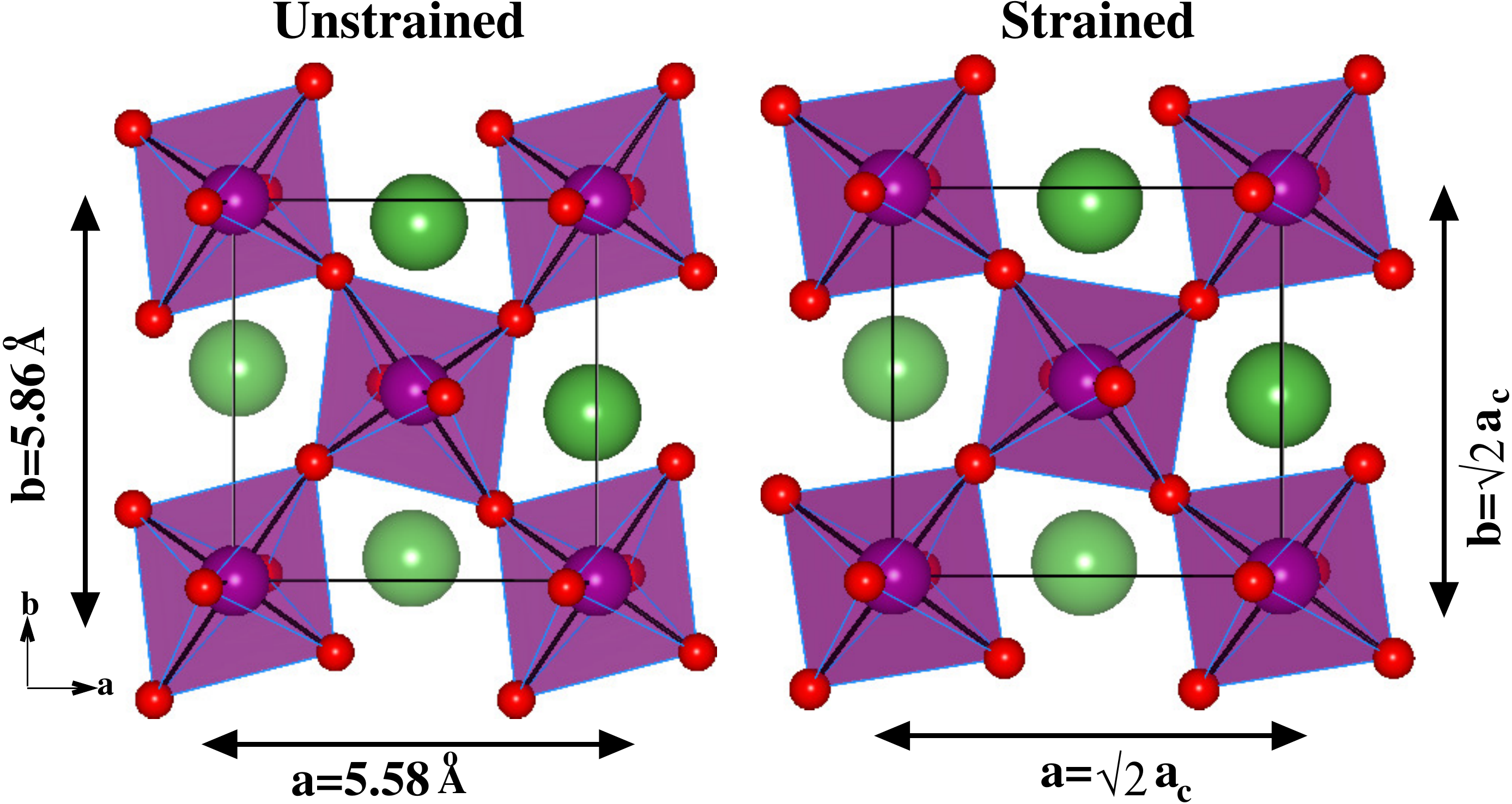}
\caption{(Color online) The structure of unstrained (left panel) and epitaxially strained LMO to square substrate (right panel) viewed
  along the crystallographic $c$-axis. Mn
 atoms (medium magenta balls) in octahedral coordination of O (small red balls) atoms share corners, while La atoms (large green balls) sit in voids.}
\end{figure}

\begin{figure}
\centering
\includegraphics[width=\columnwidth]{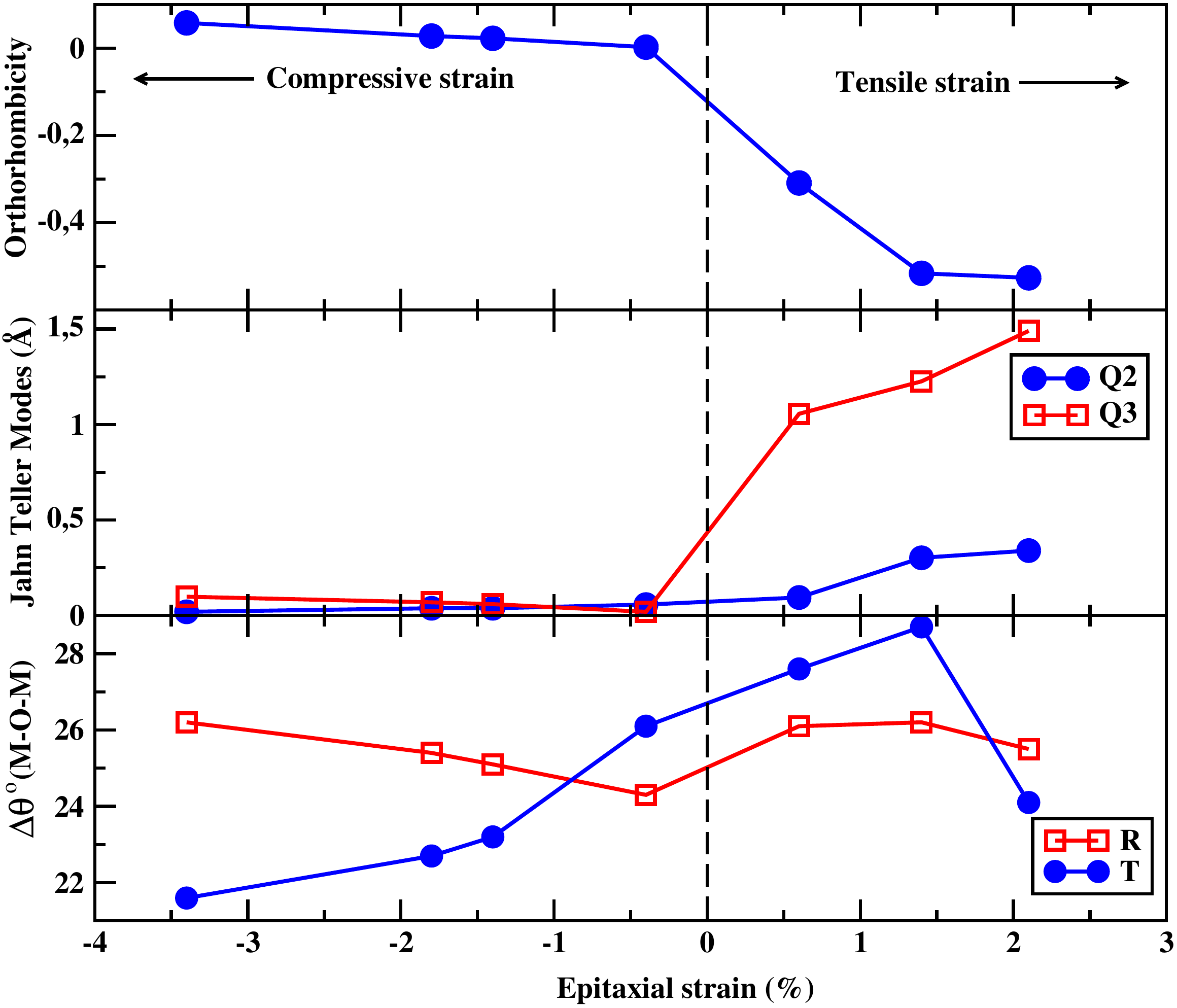}
\caption{(Color online) The structural parameters, namely orthorhombicity, JT distortions (Q2 and Q3), rotation (R)-tilt (T) of
  MnO$_6$ octahedra in the epitaxially strained bulk LMO structure.}
\end{figure}

Fig. 1 shows
the structure of unstrained and strained LMO viewed along the $c$ direction. For strained calculations,
we considered a range of strain values, varied over both compressive and tensile strain, from -3.4 $\%$ to +2.1 $\%$.
The in-plane strain produced by coherent matching of LMO to a square-lattice substrate with lattice parameter $a_c$,
was quantified as $\frac{a_c-a_0}{a_0}$ with $a_0=3.976\AA$, the cube root of the computed volume per formula unit of the
relaxed $Pbnm$ structure of unstrained LMO in its lowest energy antiferromagnetic (AFM) ground state of A-type. Fig. 2 shows the structural characteristics of the strained LMO fit
to square substrate. We find a rather strong influence of straining on the structural parameters of LMO. In particular
for tensile strain, the orthorhombicity of the $a$ = $b$ structure, defined as $c$/$\sqrt{2}a$-1 gets strongly suppressed
upon matching to a square substrate, making the structures nearly cubic (cf top panel, Fig. 2). This also gets reflected
in the JT distortion modes, defined
as $Q_2=\frac{1}{\sqrt{2}}(X_1-X_4-Y_2+Y_5), Q_3=\frac{1}{\sqrt{6}}(2Z_3-2Z_6-X_1+X_4-Y_2+Y_5)$, with ${X_i,Y_i,Z_i}$ defining
oxygen coordinates of the MnO$_6$ octahedra (see Fig. 1. in Ref.\onlinecite{spaldin-ch6}). Fig. 2 (middle panel) shows that the the JT distortion
essentially vanishes for the compressive strain. Focusing on the specific case of an STO substrate, which corresponds
to  a $a_c$ value of 3.905 $\AA$ and compressive strain of -1.8 $\%$, we find magnitudes of Q2 and Q3
to be 0.037 $\AA$, 0.068$\AA$, respectively. The tilt and rotations are also seen to be strongly influenced by the epitaxial strain
of the square substrates (cf bottom panel, Fig. 2). Our results are in good agreement with that obtained
by Lee et. al.\cite{spaldin-ch6} using a similar approach.

In the following we focus on the 1.8 $\%$ compressive strained LMO, as it corresponds to an STO substrate, and discuss
its  magnetic and electronic behavior within the framework of three approaches: DFT+$U$, hybrid and DFT+DMFT.

\begin{figure}
\centering
\includegraphics[width=1.05\columnwidth]{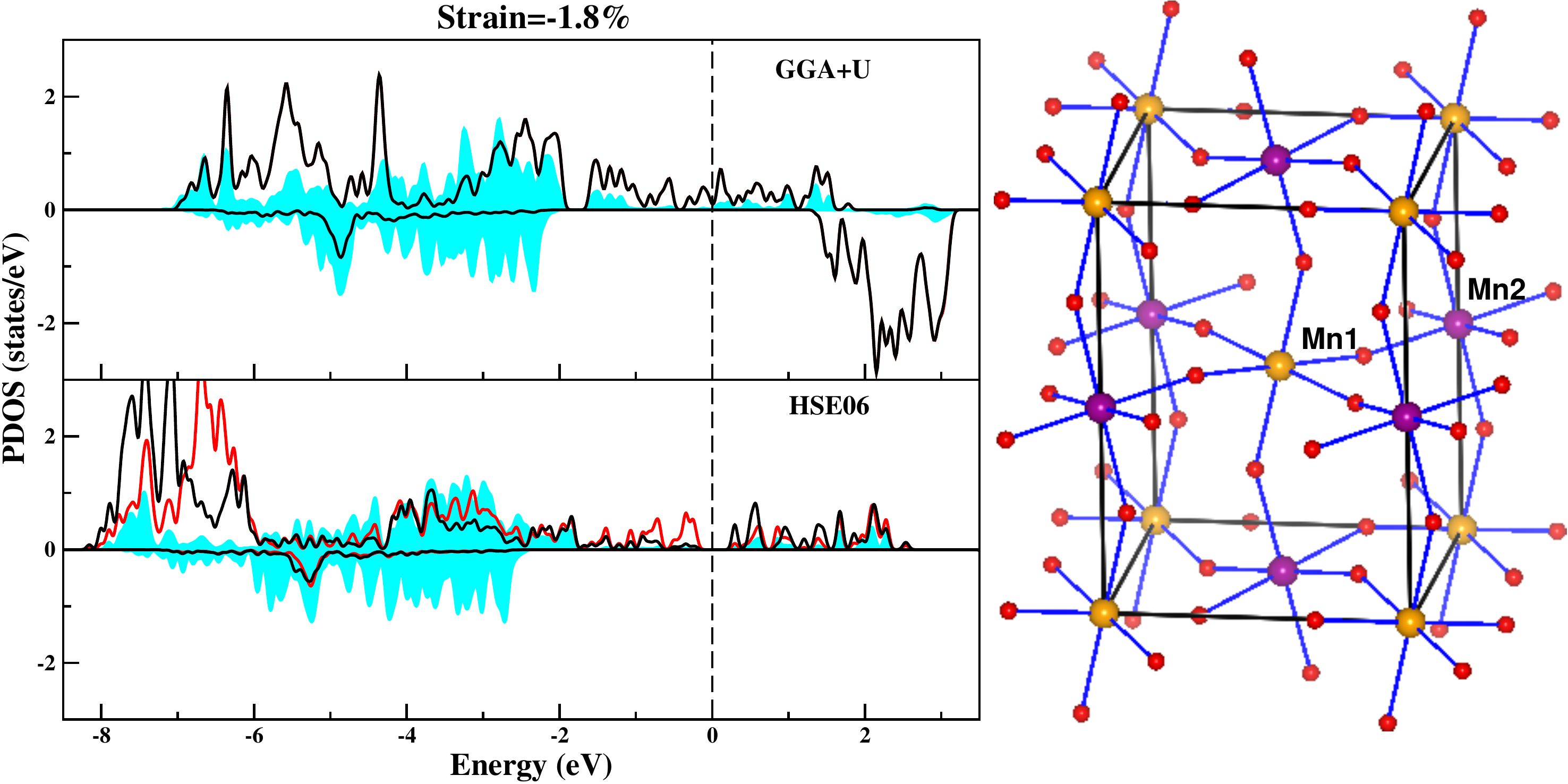}
\caption{(Color online) Left panel: Projected density of states (PDOS) for LMO strained to an STO substrate, as calculated with DFT+$U$ (top) and hybrid
  functional (bottom). The DOS projected onto the Mn $d$ states belonging to class 1 (Mn1) and class 2 (Mn2) (see text and right panel) are
  shown as black and red/grey lines, respectively; that onto O $p$ are shown as cyan/grey shaded area. The zero of energy is set to Fermi
  energy. Right panel:  The four Mn atoms in the unit cell which are structurally equivalent, however electronically
  form two classes within hybrid functional calculations, represented by Mn1 (yellow) and Mn2 (violet). }
\end{figure}

\subsubsection{Electronic and Magnetic Structure: DFT+$U$}

We calculated total energies of unstrained and strained LMO considering ferromagnetic (FM) and different antiferromagnetic
alignment of Mn spins: A-AFM, C-AFM, and G-AFM.  Here, A-AFM refers to ferromagnetic planes coupled in an antiferromagnetic manner, C-AFM refers
to antiferromagnetically arranged planes coupled ferromagnetically and G-AFM refers to antiferromagnetic alignment of Mn spins both
within the plane and in the out of plane direction. For each of the magnetic arrangements, the structure was relaxed in order to
take into account the strong influence of the structure on magnetism and vice-versa. The DFT+$U$ calculation resulted in the A-AFM insulating solution
as lowest energy solution for the unstrained LMO, in agreement with previous studies.\cite{spaldin-ch6,hou-ch6} The energy difference 
of A-AFM from FM was estimated to be small (9.01 meV/f.u), in agreement with findings by Lee et al.\cite{spaldin-ch6} suggesting the system
to be close to ferromagnetic instability which may be stabilized by external perturbation such as strain. The calculated direct band gap
of 1.3 eV, estimated from band structure, is in agreement with that of Lee et. al.\cite{spaldin-ch6} of 1.1 eV and Hou et
al.\cite{hou-ch6} of 1.2 eV, which somewhat underestimates the experimental estimate.\cite{exptl-lmo-ch6} The estimated in-plane
and out-of-plane magnetic exchanges of 2.19 meV and -1.14 meV were also found to be in good agreement with experimental estimates
of 1.85 meV and -1.1 meV, respectively.

Moving to strained LMO, we find the FM state to be stabilized by a large energy difference of about 175 meV/f.u over the A-AFM state. The suppression
of the JT distortion, which kills the orbital polarization has been argued to be responsible for the observed FM behavior.\cite{spaldin-ch6}

The left, top panel of Fig. 3 shows the DFT+$U$ density of states (DOS), projected to $d$ states of four Mn atoms in the unit cell, and the O $p$ states.
As is seen, a DFT+$U$ treatment of the problem results in a ferromagnetic half -metallic state, with empty Mn $d$ states in
the minority spin channel and filled Mn $t_{2g}$ states in the majority spin channel, while partially filled Mn $e_g$ states
hybridize with O $p$ states and cross the Fermi level. This is in agreement with the finding by
Lee et. al.\cite{spaldin-ch6} The stability of half-metallic DFT+$U$ has been checked varying the $U$ value.
Even choice of a very high value of Hubbard $U$ ($U$= 8eV), keeps the solution half-metallic. Interestingly, the solution
is found to be half-metallic even in GGA calculation setting $U$= 0. This is driven by the fact that the suppression
of the JT distortions  removes the orbital polarization, resulting in nearly
degenerate $e_g$ states, which together with  the large $e_g$ bandwidth promotes a FM spin alignment. The DFT+$U$ method which is designed to make the
configurations with larger magnetization more favorable is not effective here in a manifold of nearly degenerate bands involving only
one spin channel, though the double counting correction remains operative. The calculated moment at Mn site turned out to be 3.86 $\mu_B$
with an average moment of 0.07 $\mu_B$ arising due to the covalency effect, setting the net moment in the cell to be integer value of
4 $\mu_B$ in accordance with its half-metallic character.

\begin{figure}[tb!]
\begin{center}
\includegraphics[width=\columnwidth]{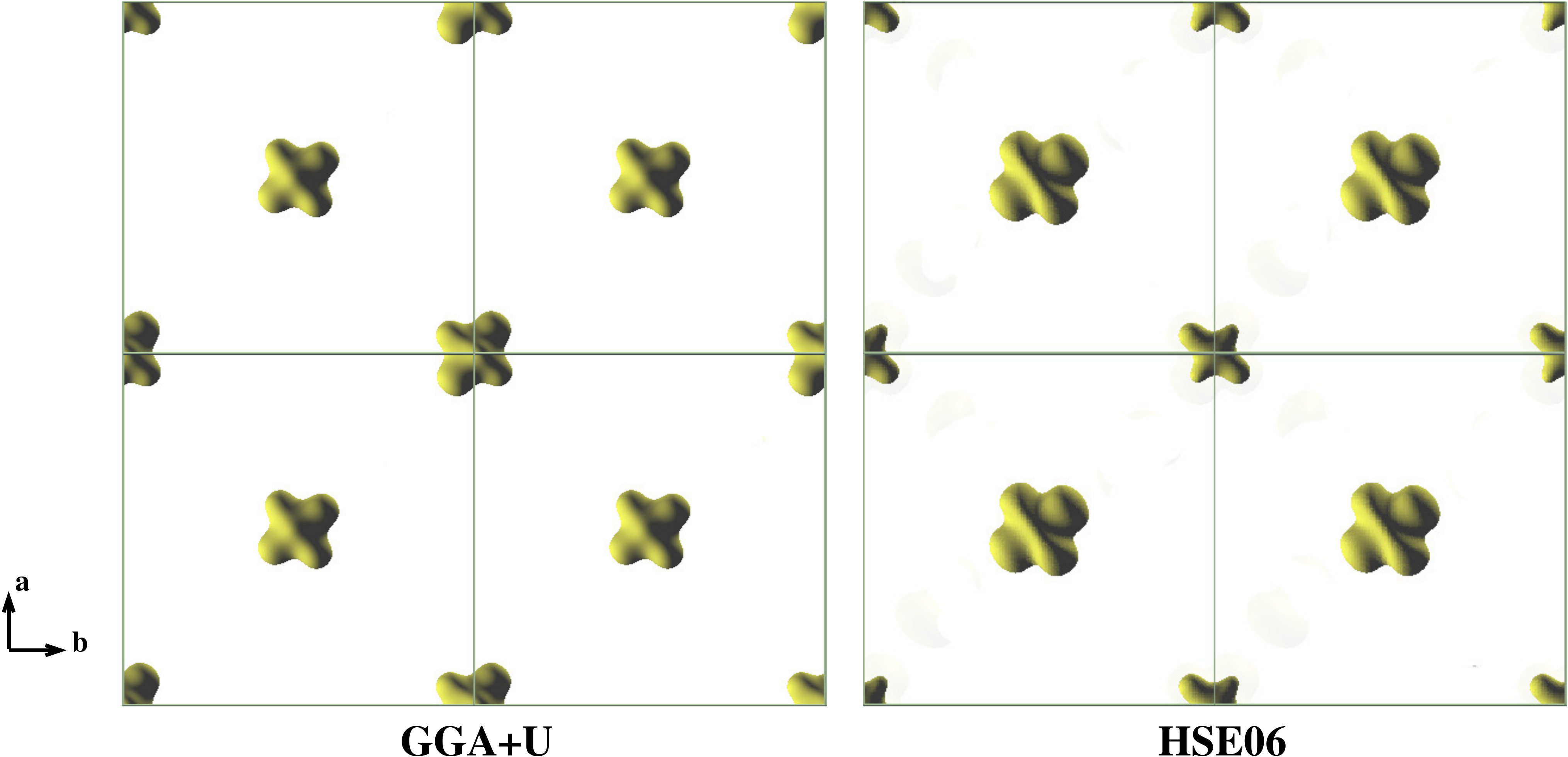}
\caption{(Color online) Charge density of LMO epitaxially strained STO, calculated within DFT+$U$ (left panel) and hybrid functional (right panel), viewed
  along $c$-axis. Shown are plots for 2 $\times$ 2 super cell for better visualization. The isosurface value is set at 0.015
  e$^{-}$/\AA$^{3}$.}
\end{center}
\end{figure}

\subsubsection{Electronic and Magnetic Structure: Hybrid}

Application of a hybrid functional with choice of $\alpha$ = 0.25 on unstrained bulk LMO, also correctly reproduced the experimentally observed
A-AFM insulating state, with a smaller energy difference of A-AFM from FM ($\approx$ 5 meV/f.u) compared to that obtained in DFT+$U$.
Our HSE06 calculation using 25\% HF exchange gave a direct band gap of 1.72eV, which is a better agreement with experimental results than
DFT+$U$, a fact mentioned already by Munoz et al.\cite{munoz-ch6} The estimates of magnetic exchanges within hybrid calculation were also
found to be reasonable with values of 2.46 meV and -0.60 meV for in-plane and out-of-plane exchanges.

However, qualitatively different result are obtained compared to DFT+$U$ when the HSE06 calculations are carried out for strained LMO.
Although FM state is found to be stabilized also in hybrid calculation by an energy difference of about 90 meV/f.u over A-AFM,
this FM state is found to be insulating, as opposed to being metallic in DFT+$U$ calculation. The DOS,
as given by hybrid functional calculation, is presented in the left, bottom panel of Fig. 3. 

The electronic structure
is found to be markedly different from that of DFT+$U$ on several counts. 
Firstly, the hybrid DOS shows a gap of about 0.3 eV,
as opposed to zero gap in case of DFT+$U$. Secondly, a substantial redistribution of spectral weights of O $p$ states is observed
in hybrid DOS, when compared to DFT+$U$. This shifts the oxygen contribution away from Fermi level. Thirdly and most
remarkably, the contributions of Mn atoms turn out to be inequivalent, with two of the four Mn atoms in cell (grouped as Mn1 in
the figure shown in right panel) occupying the body center and corners of the cell forming one class and the rest (grouped as Mn2 in
the figure shown in right panel) forming another class. The Mn2 states are found to more occupied compared to Mn1 states, suggesting
a charge disproportionation between the two. To demonstrate this we calculated the Bader charge of the Mn atoms, which is
considered to be a good approximation to the total electronic charge of an atom.  The Bader charge calculated for Mn1 and Mn2 showed
a difference of 0.07 e$^{-}$ with a difference of magnetic moment of 0.16 $\mu_B$; specifically Mn1 atoms have a charge  $4-\delta$ and Mn2
atoms have $4+\delta$. 

This is further supported by the plot of charge density in an energy window from -0.5eV below
Fermi energy up to the Fermi energy, shown in Fig. 4. While the  DFT+$U$ charge density distribution supports the ferro-orbital ordering
of type $d_{3z^{2}-r^{2}}$ + $d_{x^{2}-y^{2}}$ with no charge disproportionation between Mn atoms, as found by Hou et. al.\cite{hou-ch6},
a strong charge imbalance is noticed between Mn1 and Mn2 in the charge density plot Fig.~4, based on the hybrid functional  calculation. 

We next make an attempt to  understand this charge disproportionation,  purely driven by electronic mechanism
as the Mn atoms are structurally equivalent within the orthorhombic symmetry of the structure. We note that in CaFeO$_3$ Fe ion is in its
$d^4$ state, as is Mn in LMO. Thus Mn in LaMnO$_3$ and Fe in CaFeO$_3$ are isoelectronic i.e. both are in a $d^4$, $t_{2g}^3$ $e_g^1$ state.
For Fe $d^4$, in spite of the orbital-degeneracy, the $t_{2g}^3$ $e_g^1$  configuration remains free from Jahn-Teller
instabilities. They rather show charge-disproportionation transitions.\cite{cd-ref1-ch6}
This contrast has been argued by Whangbo et. al.\cite{cd-ref2-ch6} as following: charge disproportionation is favored over JT distortion
in CaFeO$_3$ because the covalent character is strong in the Fe-O bond, while the opposite is true for LaMnO$_3$ with
weaker covalency in the Mn-O bond.

 Putting LMO on STO, causes 1.8\% compressive strain on LMO, which in turn increases the Mn-O
covalency thereby favoring the propensity to charge disproportionation over JT distortion. Allowing the lattice to react to this
electronic instability, lowers the symmetry, making Mn1 and Mn2 structurally inequivalent in a monoclinic space group, as
found in the work by Hou et. al.\cite{hou-ch6} The calculated total energy of the orthorhombic and monoclinic structures
of strained LMO, show the monoclinic structure to be favored by a large energy gain of more than 100 meV/f.u, while DFT+$U$
calculations show only a marginal gain of 6 meV/f.u., as predicted by Hou et. al.\cite{hou-ch6}. The heterostructure geometry
though is expected to disallow the symmetry-lowering lattice instability. This interesting aspect calls for
further investigation.

\subsubsection{Electronic and Magnetic Structure: DFT+DMFT}
\begin{figure}[tb]
\includegraphics[width=8.6cm]{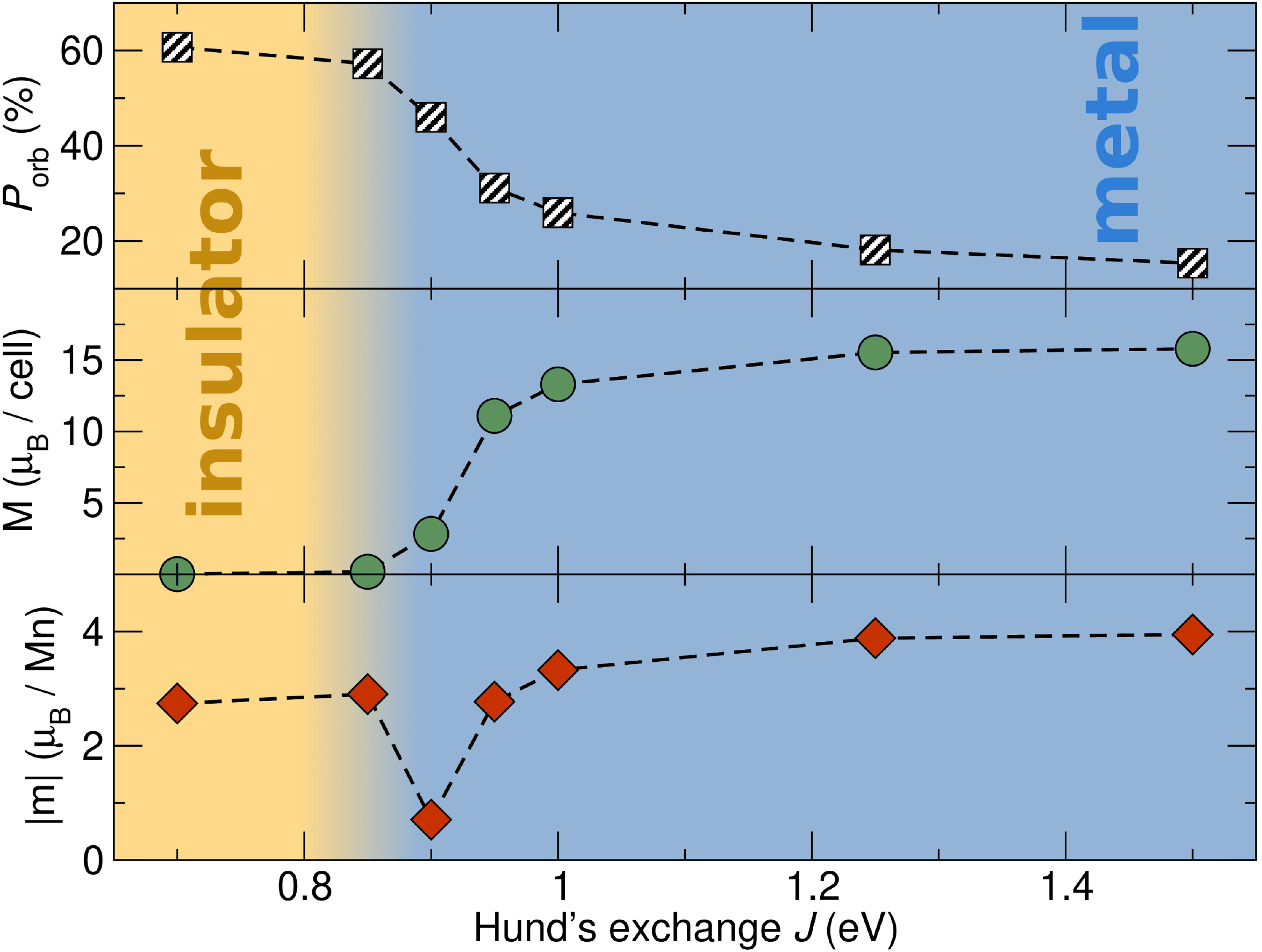}
\caption{\label{fig:dmft}(Color online) DFT+DMFT results for strained LMO at 193\,K ($\beta$\,=\,60\,eV$^{-1}$) as a function of the Hund's
exchange $J$. The inter-orbital repulsion $U'(=\!U\!-\!2J)$ is fixed to 3.6\,eV. Top, middle, and bottom panels show, respectively, the orbital polarization $P_{\text{orb}} = \left|\left(n_{3z^2-r^2}-n_{x^2-y^2}\right)/\left(n_{3z^2-r^2}+n_{x^2-y^2}\right)\right|\cdot{}100\%$, the total magnetic moment $M$ (in $\mu_{\text{B}}$) per cell, and the absolute value of the ordered moment $|m|$ (in $\mu_{\text{B}}$) averaged over the four Mn sites.}
\end{figure}

The inability to capture the experimentally observed ferromagnetic insulating
state in DFT+$U$ may also stem from a lack of dynamical effects in this
computational method.  To investigate this possibility, we performed DFT+DMFT
calculations. This way we include local dynamical correlations  between Mn $d$-electron, as well as thermal fluctuations. It is to be contrasted to  the hybrid functional which instead improves upon  the exchange, which  is a predominately non-local between the O $p$ and Mn $d$ orbitals.

For unstrained bulk LMO, we capture in DFT+DMFT  (at $U$\,=\,5\,eV, $J$\,=\,0.7\,eV, $\beta$\,=\,40\,eV$^{-1}$, not shown) the correct  room temperature
paramagnetic phase with an almost complete orbital polarization (97\%)  of the $x^2-y^2$ and $z^2$-orbitals, arguably larger than in experiment~\cite{murakami98} (please note however that our orbitals have a sizable O $p$ admixture and are defined with respect to a local [tilted] coordinate system).
As for the bulk magnetic phase occurring at lower temperatures, DFT+DMFT predicts an antiferromagnetic ground state, albeit of   G-type rather than the experimentally observed A-type. One possible reason for this discrepancy might be Hund's exchange
on ligand site being neglected.
Also the  N\'eel temperature is somewhat larger than the  experimental $T_{\text{N}}$\,=\,140\,K, as to be expected for a mean-field theory in space which neglects  non-local spin fluctuations.

Let us now turn to  our  DFT+DMFT calculations for strained LMO, with the main results shown in Fig.~5.
At         $J$\,=\,0.7\,eV, we find an antiferromagnetic insulator with a net magnetization $M=0$ and a sizable  orbital polarization. This is qualitatively similar as in the unstrained bulk, but quantitatively the strain reduces the Jahn-Teller distortion which in turn reduces the orbital polarization. If we increase $J$ slightly, there is a phase transition to a ferromagnetic metallic phase at   $J$\,=\,0.9\,eV, signaled by a net magnetization $M$ in Fig.~5. A further increase of Hund's exchange leads to the growth of the 
local moments as well as the total magnetization, accompanied by a drastic
decrease of the orbital polarization. But the ferromagnetic phase remains metallic.\cite{footnote}
Note that a similar increase of the Hund's exchange to   $J$\,=\,1\,eV
for unstrained LMO (not shown) also destroys the antiferromagnetic phase, but neither induces a ferromagnetic phase nor significantly reduces the orbital polarization up to   $J$\,=\,2\,eV.

We thus conclude that our DFT+DMFT calculations based on a five-orbital $d$-only
model yield a ferromagnet at reasonable values of $J$, but not a ferromagnetic insulator within a reasonable range of
interaction parameters.  Local, dynamic correlations within the Mn $d$ manifold are thus not enough to stabilize a ferromagnetic insulating phase.

\subsection{LMO/STO Superlattice}
\label{Sec:superlattice}

\begin{figure}[tb]
\includegraphics[width=8.6cm]{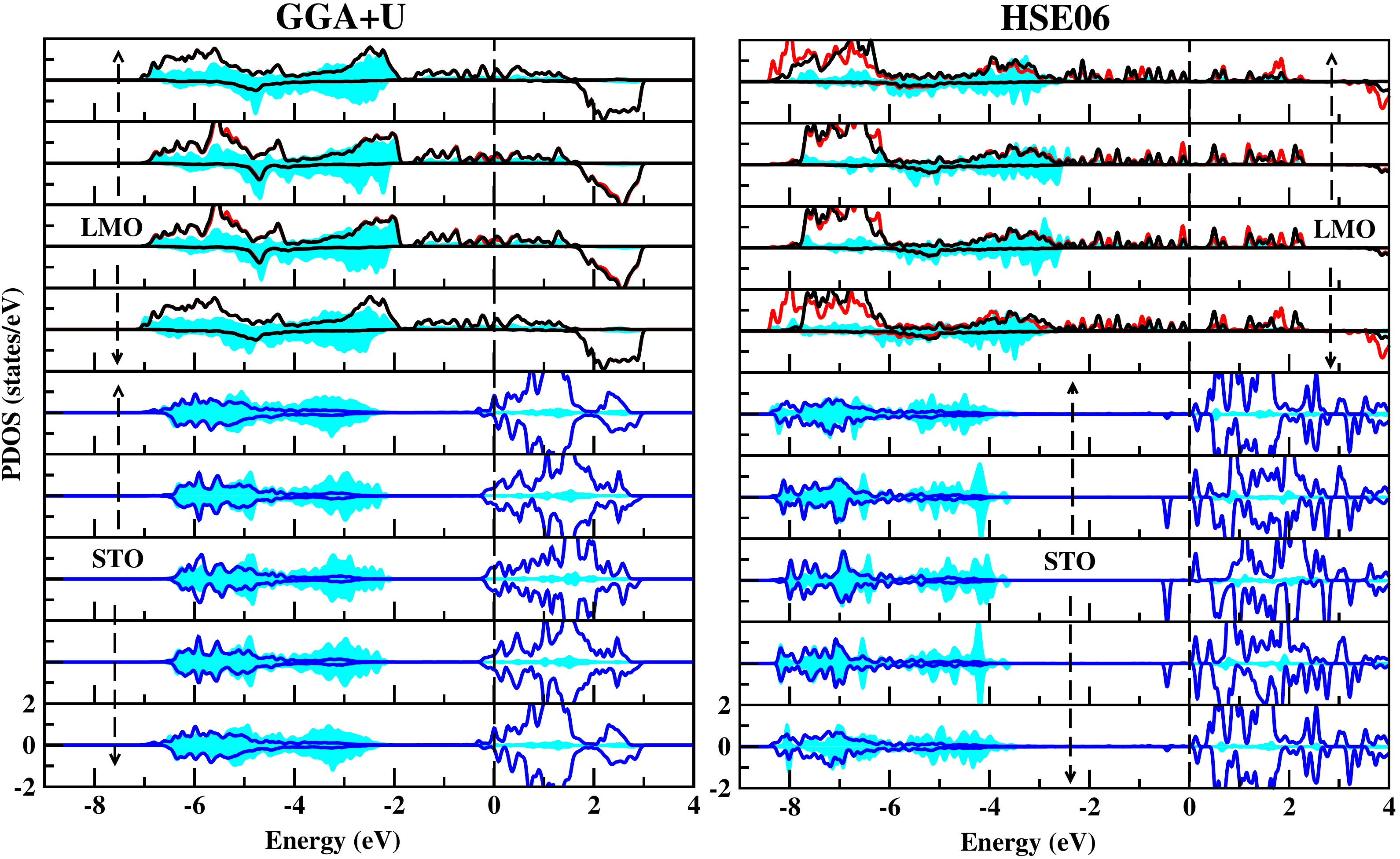}
\caption{(Color online) Layer projected DOS for a  (LMO)$_{4.5}$/STO$_{4.5}$ superlattice with states projected to Mn1 $d$ (black line), Mn2 $d$ (red/grey line), and O $p$ (cyan/grey shaded) states. The zero of energy is set to the Fermi energy. The left panel shows the
 DOS calculated in DFT+$U$ while the right panel shows the DOS calculated in HSE06 hybrid functional.}
\end{figure}

We next investigate the electronic structure of LMO on STO in experimental set-up, i.e., we consider an actual heterostructure including the STO layers.
In  addition to the square epitaxial strain
of the “strained-bulk” structure, this in particular involves the polar discontinuity formed between LMO consisting of alternating layers of LaO and MnO$_2$
of +1 and -1 charges and STO consisting of alternating charge neutral layers of SrO and TiO$_2$. The latter would cause half a charge
to be transferred between the layers at the interface. Neither the direction or extend of this charge transfer has been clarified.
The electronic phase separation suggested in recent study,\cite{sumilan-ch6} is based on the assumption of charge being transferred
to LMO, thus doping LMO. To the best of our knowledge, this has not been verified in terms of first-principles calculations or experiment.

For the investigation of the LMO/STO interface as in the experimental situation, we considered superlattice geometry of LMO/STO, with alternate
repetition of equal thickness unit cells (4.5) of LMO and STO along the [001] direction, which creates two symmetric $n$-type
interfaces between LaO of LMO and TiO$_2$ layer of STO.\cite{banerjee-ch6,oor} Equal thicknesses were chosen since the
FM insulating state has been experimentally observed for superlattice geometries with nearly equal thickness of LMO and STO layers.\cite{oor}
The electronic and magnetic structures of the constructed superlattice geometry was investigated within charge self-consistent DFT+$U$
and hybrid functional calculations. The DFT+DMFT calculation became prohibitively expensive for such geometry.

We placed LMO in an orthorhombic geometry matching to square plane STO layers (in the [100] and [010] directions). Here a $\sqrt{2} \times \sqrt{2} \times c$ supercell of both LMO and STO was allowed to tilt and rotate. This
resulted in four Mn and Ti atoms in each MnO$_2$  and TiO$_2$ layer, respectively.
The ionic positions and $c$ lattice parameters were allowed to relax, keeping the constraint of $a$ = $b$ lattice parameters.
This set-up generates a square matched epitaxial strain of -1.8\%, as in previous discussion.
The optimized structure shows a significant decrease in Jahn-Teller distortion and modification of
tilt and rotation angles in LMO, having a similar trend as in “strained-bulk” calculation, 
while some JT distortion and tilt and rotation is introduced in the STO block
layers due to its proximity to largely distorted LMO block, very similar to that found
for GTO/STO.\cite{banerjee-ch6}

Both DFT+$U$ and hybrid calculation found FM magnetic state to be stabilized compared to three different antiferromagnetic structures, i.e., A-AFM, C-AFM and G-AFM. This again confirms the experimentally observed ferromagnetic  state, and   is in a general agreement with the ``strained-bulk'' calculations above.

The left panel of Fig. 6 shows the layer decomposed partial DOS projected to Mn $d$, Ti $d$ and O $p$ in MnO$_2$ and TiO$_2$
layers of LMO and STO block, as given in DFT+$U$. We notice the FM electronic state to be half-metallic in each MnO$_2$ layers of LMO,
as found in DFT+$U$ DOS of the ``strained-bulk'' structure. The TiO$_2$ layers are metallic as well, with the Ti $d$ states at  the interface (IF) TiO$_2$
layer being spin polarized. This suggests the polar charge to reside within the STO block. This expectation turned out to be true, with
total conduction charge in STO block to be 1 e$^{-}$, being consistent with the presence of two symmetric interfaces in the unit cell, and
a carrier density of 0:5e$^{-}$ per IF.

 The situation changes dramatically in the hybrid calculation in Fig.~6 (right), where both LMO and STO blocks are
found to be insulating. The charge disproportionation-driven opening of a band gap is observed in the LMO block, similar to that found in hybrid calculations of the ``strained-bulk'' structure. What is very interesting is that the electron gas generated due to polar catastrophe in the STO side
becomes fully spin polarized in the TiO$_2$ layers within the framework of hybrid calculation. The one   extra, spin-polarized electron induced by the two $n$-type interfaces is located at the center of the STO block, causing a gap to arise at
the Fermi level even in the STO block. This turns the entire system insulating.

\section{Summary and Discussion}
With the aim to provide an  understanding regarding the  intriguing reports of a ferromagnetic insulating state in LMO/STO heterostructures,
we study the problem within a “strained-bulk” LMO structure using three different theoretical approaches, i.e., DFT+$U$, hybrid functionals, and DFT+DMFT.
We find that epitaxial straining to a square substrate of STO results in ferromagnetic ground state in all three approaches, for
reasonable choices of parameters. This primarily results from a strong suppression of the JT distortion, which quenches the orbital polarization and hence antiferromagnetism in turn.
The electronic state of the ``strained-bulk'' structure however turned out to be different between the three approaches. DFT+$U$ and DFT+DMFT
resulted in metallic solution, supporting a double-exchange scenario to be operative. The treatment of exchange effects within a hybrid functional, on the other hand, resulted in an insulating solution. At the microscopic level, the latter is an  
electronically driven charge-disproportionation among the Mn atoms in the unit cell. We note that charge-disproportionation is favored
over JT distortion in case of strong metal-oxygen covalency as found in case of Fe $d^4$ in CaFeO$_3$,\cite{cd-ref2-ch6} making propensity
to charge disproportionation in the enhanced Mn-O covalency of the strained LMO structure to be a plausible scenario.

We further investigated the case of LMO/STO superlattice structure with comparable thicknesses of LMO and STO within the schemes of DFT+$U$
and hybrid calculations. As for the ``strained-bulk'' structure, it was found that DFT+$U$ and hybrid calculations yielded FM metallic
and FM insulating states, respectively. In the hybrid functional calculation, the origin of the insulating state in LMO and STO blocks is different: the  LMO layers again undergo an electronically driven charge disproportionation, while the latter develop a complete spin polarization and localization of the constituent polar charge.

Our study employing three different, complementary theoretical tools provides an exhaustive
description of the problem. Nevertheless, each of the employed methods has
limitations that deserve a discussion.  First of all, our analysis is
restricted to stoichiometric LMO. Possible anionic or cationic defects may have drastic
consequences for the electronic and magnetic state, such as admixtures of   Mn$^{2+}$ or Mn$^{4+}$ or localization of charge carriers due to disorder. Modeling of such
situations however requires a better characterization of the experimental samples.  

The DFT+$U$ approach is designed to make configurations with larger
magnetization favorable by transferring charges between occupied and unoccupied
states in the two different spin channels, and thus may not be best tool to
describe the FM insulating state, when starting from a half-metallic solution of
DFT.                                          Our DFT+DMFT study is based on the Mn $d$-only model and has been carried out within
the so-called single-shot scheme.  While this incorporates the electronic correlations of the Mn $d$ orbitals, the $d$-$p$ exchange and a possible charge redistribution due to electronic correlations is neglected.  Both effects  may turn out to be important for an
accurate description of magnetic and electronic state in the presence  of strong
covalency effects. Furthermore, while local correlations are accounted for in DMFT, nonlocal correlations require cluster~\cite{maier05} or
diagrammatic~\cite{rohringer18} thereof, which are 
presently unfeasible in our case. Hybrid functionals incorporate a better treatment of the exchange,  predominately between O $p$ and Mn $d$ states. In our calculation this exchange eventually drives a purely electronic charge disproportionation and  opens an insulating gap.

This charge disproportionation of the hybrid functional solution provides at least one route to the experimentally observed ferromagnetic insulator, while we cannot exclude that other effects such as disorder 
due to cation intermixing or oxygen vacancies also play a role.
 We conclude that both, an improvement of
the theoretical approaches as well as a better characterization of experimental
samples, are needed to have a complete understanding of these complex LMO/STO heterostructures.

\appendix*
\section{}
\begin{figure}[b]
	\includegraphics[width=0.7\columnwidth]{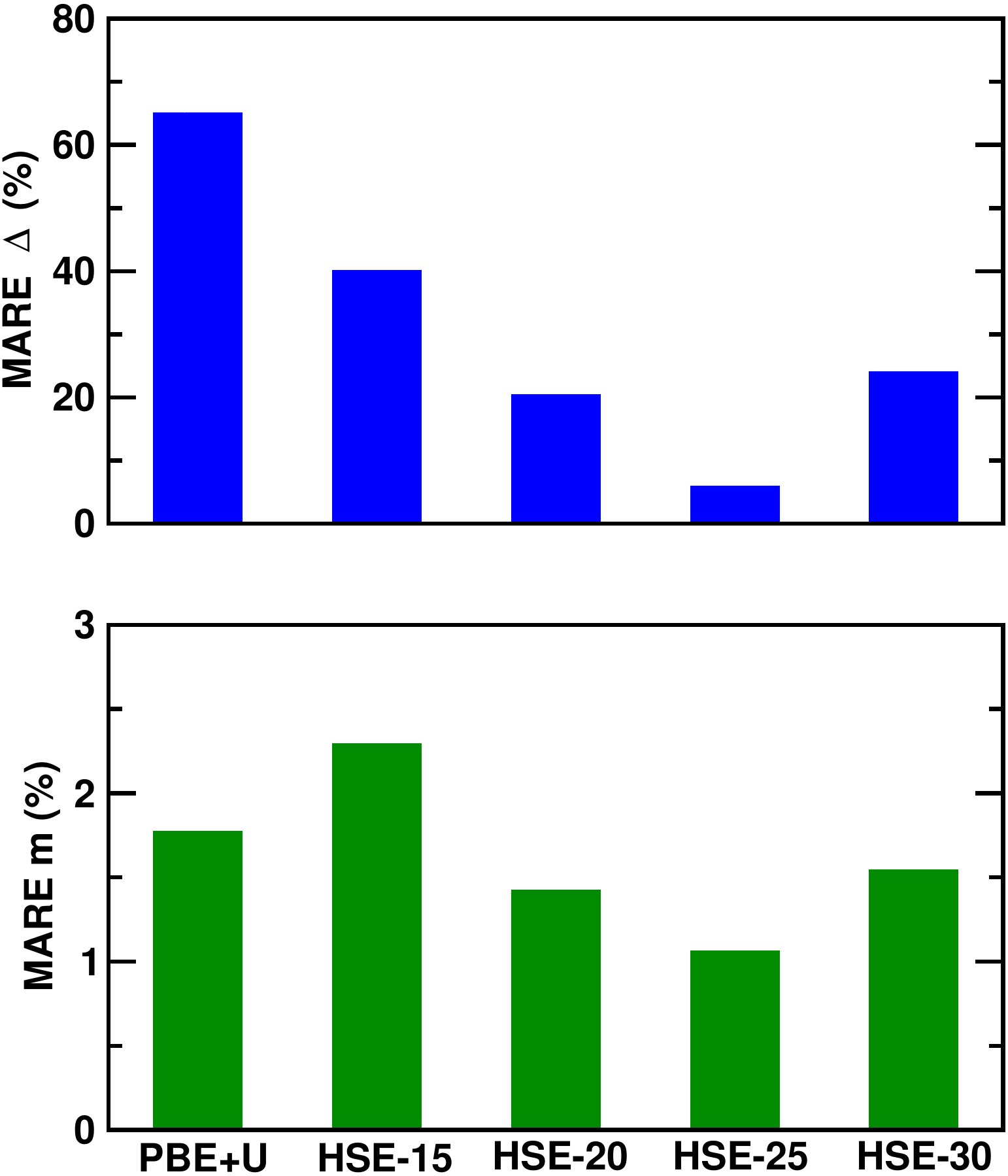}
	\caption{(Colour online) MARE (\%) values for band gaps (top) and magnetic moments (bottom) for  different mixing parameters  $\alpha$ of the HSE functional. Shown are also DFT+$U$ results.}
\end{figure}

For determining the mixing parameter $\alpha$ of the hybrid  HSE functional, we compared the calculated band gap and magnetic moment for bulk LMO with experiment. 
Experimental band gap values of 1.7eV\cite{exptl-lmo-ch6}, 1.9eV\cite{jung1} and 2.0eV\cite{jung2, kruger} have been reported; and experimental magnetic moments were been found to be 3.87$\mu_B$\cite{moussa} and 3.7$\mu_B$.\cite{elemans} Here we disregarded the outliers. 

To account for the experimental variation, we plot in 
Fig. 7 the Mean Absolute Relative Error (MARE) as mentioned in Section II.
As a percentage, it reads  $$MARE=\frac{100}{n}\sum_{i=1}^{n}\mid\frac{E_i-T}{E_i}\mid$$
where $E_i$ represents the experimental value, $n$ the number of experimental values considered, and $T$ the theoretically calculated value  for one particular functional. 

Fig. 7 shows that the MARE is minimal for $\alpha$=0.25, indicating that  $\alpha$=0.25, rather than $\alpha$=0.15,\cite{cesare} provides 
best agreement with the experimental band gap and magnetic moment if a higher cut-off for the potential is choosen, at least for LaMnO$_3$.


\begin{thebibliography}{99}
\bibitem{footnotedmft} This scheme is often employed in DFT+DMFT calculations of oxide heterostructures when charge
self-consistent calculations become prohibitively expensive~\cite{janson18a},
see Refs.~\onlinecite{janson18b, beck19X} for recent examples.
\bibitem{footnote} {Interestingly, if we now fix
	$J$\,=\,1.25\,eV, which corresponds to 20\% orbital polarization with the
	nearly saturated magnetization (Fig.~\ref{fig:dmft}), and increase the
	inter-orbital interaction $U$ from 3.6 to 4.5\,eV, the orbital polarization
	jumps to 66\%, while magnetization drops down to 1.9\,$\mu_{\text{B}}$ per cell
	(not shown).}
\end{thebibliography}
\end{document}